\documentclass[a4paper,11pt]{article}
\pdfoutput=1 

\usepackage{jinstpub} 

\usepackage[utf8]{inputenc}
\usepackage{amsmath}
\usepackage{amsfonts}
\usepackage{graphicx}
\usepackage[dvipsnames]{xcolor}
\usepackage{textgreek}
\usepackage{verbatim}
\usepackage{caption}
\usepackage{enumitem}
\usepackage{makecell}
\usepackage{tikz}
\usepackage{tikz-timing}
\usetikzlibrary{shapes.symbols, arrows, calc, shapes.gates.logic.US, intersections}
\usepackage{svg}

\usepackage{lineno}
\notoctrue

\title{RD50-MPW3: A fully monolithic digital CMOS sensor for future tracking detectors}

\author[a,1]{Patrick Sieberer \note{Corresponding authors.}}
\author[b,1]{Chenfan Zhang}
\author[a]{Thomas Bergauer}
\author[c]{Raimon Casanova Mohr}
\author[a]{Christian Irmler}
\author[b]{Nissar Karim}
\author[d]{Jose Mazorra de Cos}
\author[a]{Bernhard Pilsl}
\author[b]{Eva Vilella}

\affiliation[a]{Austrian Academy of Sciences, Institute of High Energy Physics,\\ Nikolsdorfer Gasse 18, 1050 Wien, AT}
\affiliation[b]{Department of Physics, University of Liverpool,\\Oliver Lodge Building, Oxford Street, Liverpool L69 7ZE, UK}
\affiliation[c]{Institute for High Energy Physics (IFAE), Autonomous University of Barcelona (UAB), \\ Bellaterra, 08193, Barcelona, ES}
\affiliation[d]{Instituto de Física Corpuscular (IFIC), CSIC-UV. \\ Parque Científico,
Catedrático José Beltrán, 2, 46980 Paterna (Valencia), ES}

\emailAdd{patrick.sieberer@oeaw.ac.at}
\emailAdd{chenfan@hep.ph.liv.ac.uk}

\abstract{The CERN-RD50 CMOS working group develops the RD50-MPW series of monolithic high-voltage CMOS pixel sensors for potential use in future high luminosity experiments such as the HL-LHC and FCC-hh.
In this contribution, the design of the latest prototype in this series, RD50-MPW3, is presented. An overview of its pixel matrix and digital readout periphery is given, with discussion of the new structures implemented in the chip and the problems they aim to solve. 
The main analogue and digital features of the sensor are already tested and initial laboratory characterisation of the chip is presented.
}

\keywords{CERN-RD50, RD50-MPW3, DMAPS, HVCMOS, Future Hadron Colliders}

\arxivnumber{2211.11244} 

\collaboration[c]{\\ \centering\includegraphics[height=20mm]{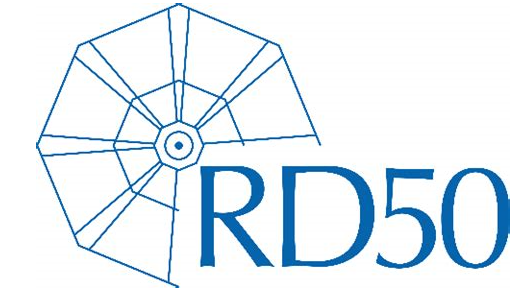}\\[6pt] on behalf of the CERN-RD50 collaboration}

\proceeding{TWEPP 2022\\
  19.–23. Sept. 2022 \\
  Bergen, Norway}

\begin{document}
\maketitle
\flushbottom
\section{Introduction}
\begin{figure}[htbp!]
    \centering
    \includegraphics[width=0.89\textwidth]{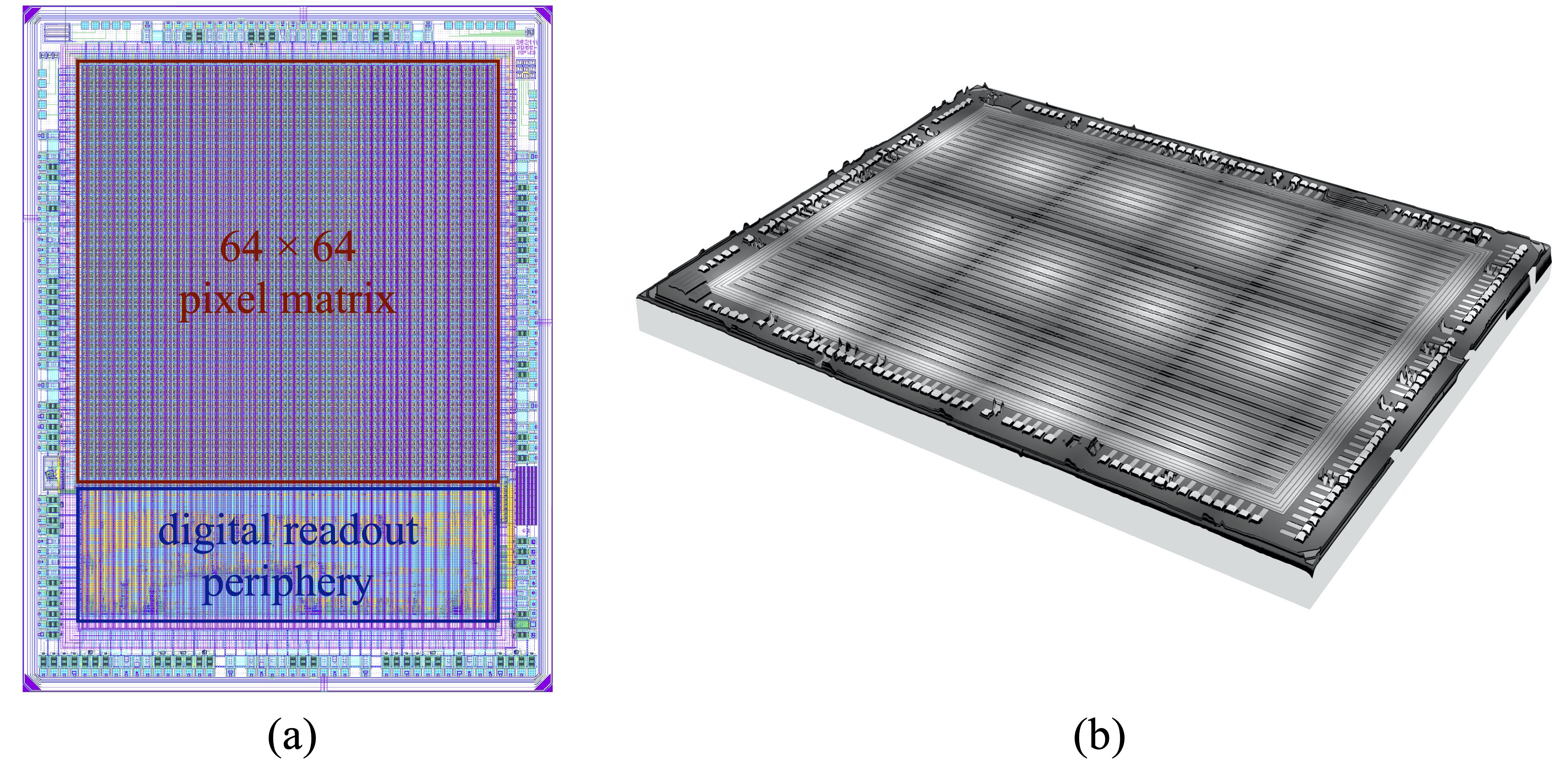}
    \caption{(a) Layout of RD50-MPW3; (b) 3D picture taken with a 3D surface metrology microscope.}
    \label{fig:chip_views}
\end{figure}

To further improve and demonstrate the performance of Depleted Monolithic Active Pixel Sensors (DMAPS),
the CERN-RD50 CMOS group has designed a series of prototype chips \textcolor{black}{using the 150~nm HV-CMOS process from LFoundry S.r.l}. Following the first two designs, RD50-MPW1 and RD50-MPW2 \cite{Eva:Vertex2019, EvaNIMA}, a further chip called RD50-MPW3 has been developed. The development of these sensors targets high granularity for precise tracking
and tolerance to harsh radiation environments ($1 \times 10^{15}~\mathrm{n_{eq}/cm^2}$ and 1~MGy) in future experiments. Figure \ref{fig:chip_views} shows the layout view and a 3D picture of the chip.

RD50-MPW3 incorporates digital readout electronics based on the column-drain architecture in its $64 \times 64$ pixel matrix, and has an optimised digital periphery for effective chip configuration and fast data transmission. This paper presents the design of the chip by describing its two main components, the pixel matrix and digital periphery, in section \ref{sec:pixel_matrix} and section \ref{sec:digital_periphery} respectively. Moreover, first results from lab measurements of RD50-MPW3 are shown at the end of the paper.


\section{Pixel Matrix}\label{sec:pixel_matrix}
The pixel matrix has both analogue and digital in-pixel electronics for processing  and sending out hit events. The design optimises in noise reduction and power distribution.

\subsection{Double-Column Architecture}
\begin{figure}[htbp!]
\centering
    \includegraphics[width=0.74\textwidth]{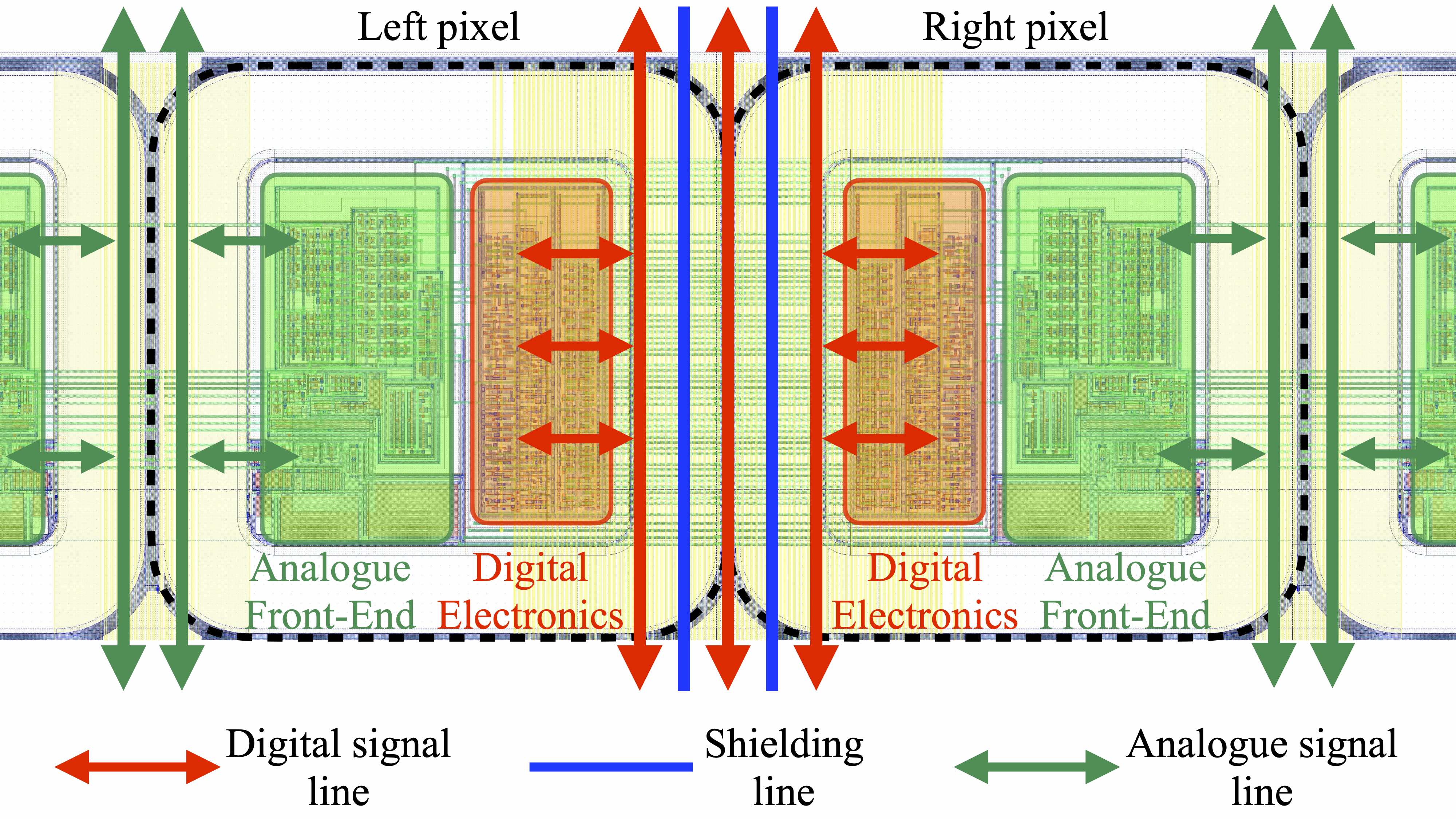}
    \caption{Two pixels in a double column. Digital signal lines are routed in the middle and shared by the two pixels. Analogue signal lines are shared between adjacent double columns.}
    \label{fig:double_column}
\end{figure}

In one of RD50-MPW3's predecessors RD50-MPW1, crosstalk noise was detected between signal lines that are routed with the minimum spacing allowed by the design rule \cite{Eva:Vertex2019}. To alleviate routing congestion and minimise coupling between metal lines, the pixel matrix of RD50-MPW3 is designed using the double-column architecture, where 64 pixel columns are organised into 32 double columns (128 pixels in each double column). Figure~\ref{fig:double_column} illustrates two pixels in a double column.

The left and right pixels are placed in a mirrored pattern, with their digital readout electronics facing each other.
Digital signal lines are routed in the middle of the double column, and therefore can be shared by the two pixels and connect to their digital electronics without crossing analogue sections. Analogue signal lines are routed between adjacent double columns and shared by their analogue electronics. In this way, the crosstalk between analogue and digital signal lines is prevented, and the number of signal lines is halved as two pixels share the same signal lines. In order to minimise the coupling noise between the metal lines that transmit high-frequency (up to 40~MHz) digital signals, grounded shielding lines are placed between each of them. \textcolor{black}{With these design efforts, simulations show that the crosstalk noise is kept at a negligible level.}

Since a voltage drop of 200~mV was measured in RD50-MPW1 along its power rail, a power grid consisting of vertical and horizontal metal lines is incorporated in the pixel matrix to achieve a balanced power distribution. In addition, a power ring is created around the pixel matrix to supply the power grid from all four sides. \textcolor{black}{Those power lines use the top metal layers and their widths are designed to be the largest set by the maximum metal density rule. As a result, the voltage drop is reduced to 10~mV according to simulation.}

\subsection{Pixel Design} \label{sec:pixelDesign}
\begin{figure}[htbp!]
\centering
    \includegraphics[width=1\textwidth]{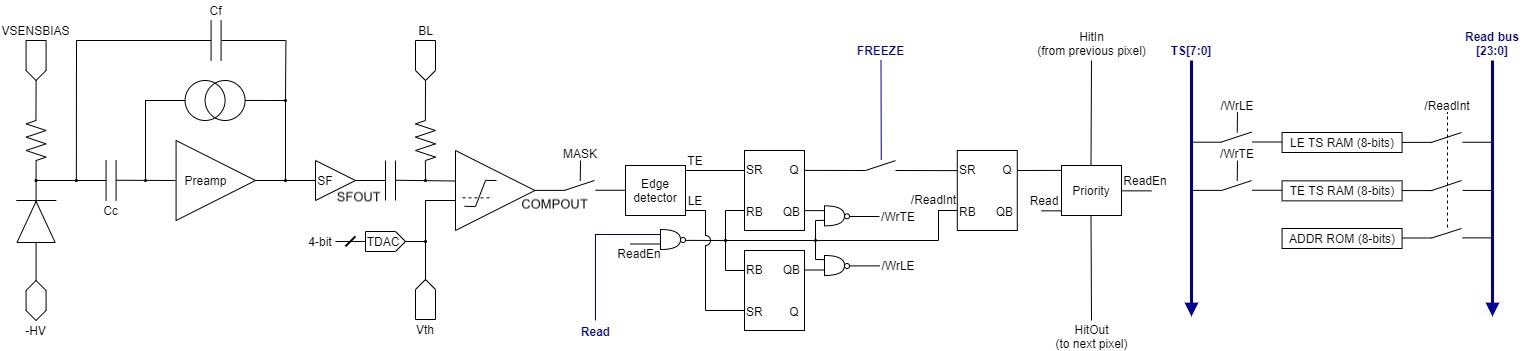}
    \caption{Simplified block diagram of in-pixel electronics, including both analogue circuits (a charge sensitive amplifier, a high-pass filter, a discriminator and a four-bit trim DAC) and digital circuits (readout control logic, an eight-bit address ROM and two eight-bit timestamp RAMs).}
    \label{fig:in_pixel_electronics}
\end{figure}

The pixel size in RD50-MPW3 is $\mathrm{62 \times 62~\mu m^2}$. Its pixel sensing diode for charge collection is in the same structure as in its predecessors \cite{twepp2019_Chenfan}. \textcolor{black}{The diode is implemented by means of a p-substrate/deep n-well junction. To expand the diode depletion region, the p-substrate is biased at a large reverse voltage using a topside p-well implant.}
To further minimise the crosstalk between analogue and digital domains, they are embedded into separate deep p-type wells.
Figure \ref{fig:in_pixel_electronics} shows a simplified block diagram of in-pixel electronics.

The analogue front-end electronics is based on that developed for RD50-MPW2 pixels, which is designed to process particle hits at high rates \cite{twepp2019_Chenfan} and has been evaluated in laboratory \cite{hernandezLatestDepletedCMOS2021, hitiCharacterisationAnalogueFront2021} and at beam facilities \cite{Sam:38thRD50, tipp}. \textcolor{black}{Its response time to an $8~\mathrm{ke^-}$ charge signal is less than 100~ns when the chip was bias at $\mathrm{-60~V}$.} 

The digital readout electronics implements improvements on that used in RD50-MPW1 \cite{Eva:Vertex2019}. It adds a logic to prevent noisy pixels from writing data bus by enabling the signal MASK \textcolor{black}{as seen in figure \ref{fig:in_pixel_electronics}}. A FREEZE signal is used to pause processing new hits until the readout of a previous event in a double column has finished. A more compact priority circuitry, which controls the right of all pixels in a double column to access the data bus, is designed based on an OR gate chain. Each pixel contains eight SRAM cells (not shown in Figure \ref{fig:in_pixel_electronics}) used for configuration, which can be written through a single-bit D flip-flop and eight selection signals. The D flip-flops of all pixels in a double column are connected to one another and form a 128-bit shift register.

\section{Digital Periphery}\label{sec:digital_periphery}
The main purpose of the digital periphery is to configure, control and readout the pixel matrix. It has been developed with digital tools and integrated into the chip using an analogue-on-top flow.
\subsection{Pixel Matrix Configuration and Readout}
\begin{figure}[!ht]
    \centering
    \includegraphics[width=0.5\textwidth]{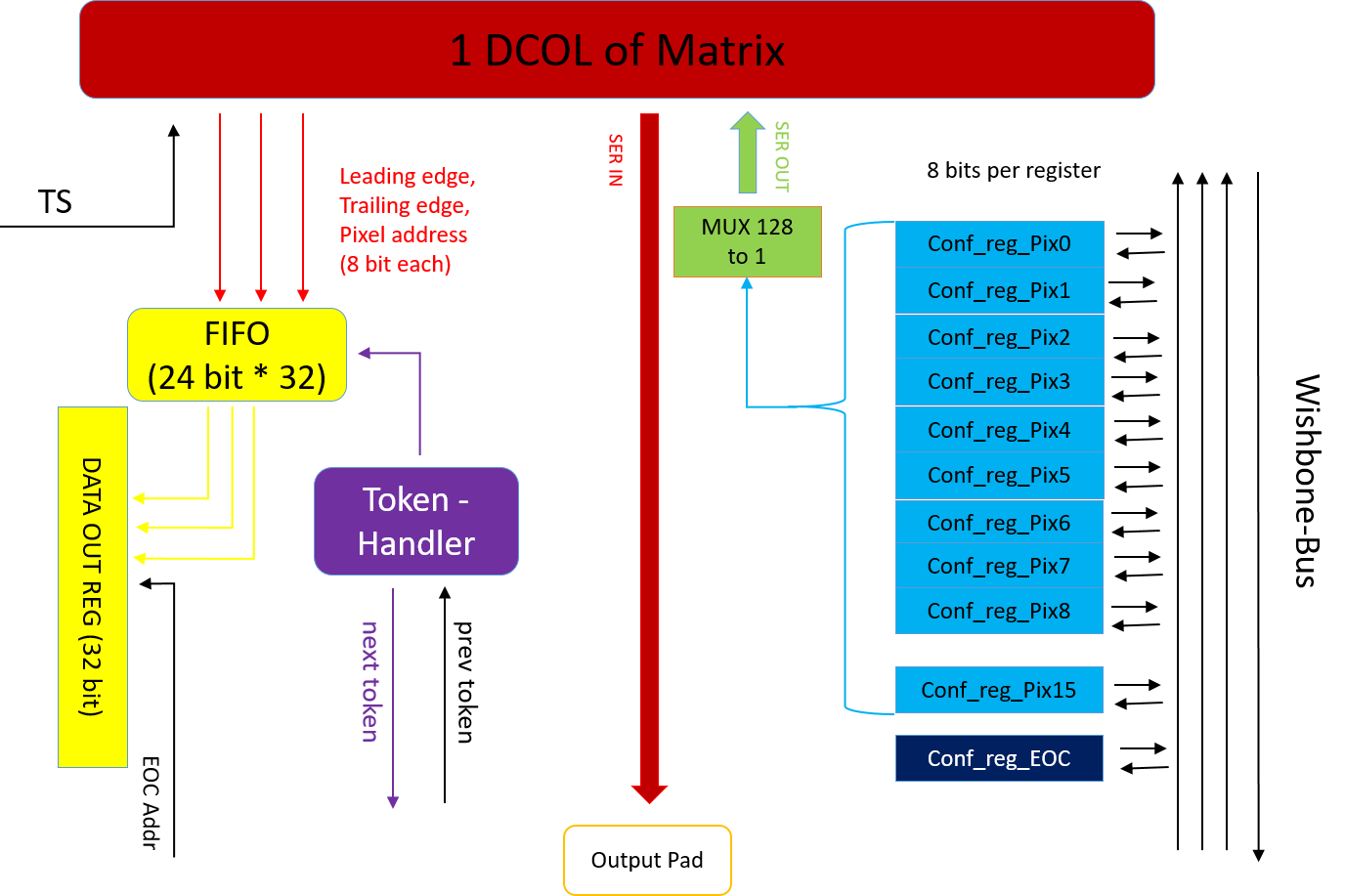}
    \caption{The left part of the schematic shows the blocks needed for the readout while the right side depicts blocks needed for configuration.}
    \label{fig:EOCschematic}
\end{figure}

Each double column of the pixel matrix has its own dedicated end-of-column (EOC) which implements the readout and configuration of a single double column. 
16 eight-bit registers are implemented in the EOC to store 128 bits as shown in figure \ref{fig:EOCschematic}. The content of these registers is bit-wise shifted into the double column by a 128 to 1 multiplexer and thereafter pushed into the in-pixel SRAM cells.
The 16 EOC registers are connected to a wishbone bus, which is used to write and read these registers by the off-chip electronics via an I2C interface. 
The configuration register defines the in-pixel configuration SRAM cell and is moreover used to control the readout of the double column which is depicted in figure \ref{fig:EOCschematic} as well. \textcolor{black}{Two different readout modes are available. One readout mode uses a simple priority logic, where pixels with a higher address are prioritized. As this introduces a bias towards pixels with high addresses, a second mode has been implemented, called the FREEZE mode, where the readout of each pixel is frozen until the whole double column has been read out. This mode uses the FREEZE signal as mentioned in section \ref{sec:pixelDesign}.}
Timestamps for the leading and trailing edges, and the pixel address, consisting of eight-bits each are transmitted from the matrix to the EOC. This information is stored in a 32-words deep FIFO for de-randomization of data from a single double column. 

The readout of the 32 EOCs is controlled by a token handler, which connects the individual EOCs one after the other with a 32-bit buffer register to temporarily store the hit information together with an eight-bit EOC address. In the next step the data of the buffer register is pushed into a transmission FIFO, while the token is given to the next EOC. Finally, a transmission unit is responsible for framing, encoding and serialization of data.
\textcolor{black}{Using a 640\,MHz LVDS line for readout, a pixel hit rate of about 16\,MHz over the full matrix can be handled.}
A detailed description of the transmission unit including timing has been published in \cite{vci}.

\subsection{Other Peripheral Features}
The periphery takes care of few more features which are listed here for completeness but not explained in detail. A common timestamp generator creates an internal timestamp which is used by the pixels to determine the leading edge and the trailing edge of the signal. It is designed to nominally run at 40\,MHz. The clock and reset unit divides a fast external clock, which is expected to be 640\,MHz, to the internally used 40\,MHz clock. This unit, moreover, takes care of two different reset signals and synchronization of the clocks as well. An I2C to wishbone module is implemented in order to convert externally used I2C commands to the internally used wishbone commands in order to allow communication with the chip over a standardized interface. 
Many analogue and digital debug outputs and external control signals are implemented for precise control and to monitor certain features of the chip. Those are not needed for normal operation of the chip, but useful for debugging purposes or direct measurement of analogue signals.

\section{Initial Measurements}
\begin{figure}[!ht]
    \centering
    \includegraphics[width=0.55\textwidth]{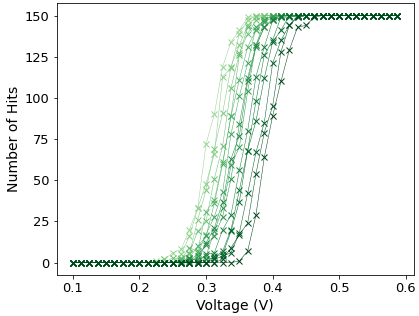}
    \caption{S-curves of a pixel for different threshold values which is set by the trim DAC. 150 hits are injected for each measurement point. Darker coloured lines corresponds to higher thresholds.}
    \label{fig:SCurve}
\end{figure}

Initial laboratory measurements have been conducted to characterise the chip performance. Figure \ref{fig:SCurve} shows S-curve measurements of a single pixel with different trim DAC settings. The in-pixel trim DACs are able to fine tune the threshold voltage of each pixel.
The expected behaviour that the pixel fires at a higher voltage, given a higher threshold, can be seen in figure \ref{fig:SCurve}. \textcolor{black}{A global threshold of 300~mV above baseline has been used for this initial study. According to simulations with RD50-MPW2, this threshold corresponds to around 5500 electrons. It is expected from results of previous submissions that the threshold can be lowered to around 50-100~mV. Measurements to confirm this value for RD50-MPW3 are on-going.}

This measurement has been conducted using the full digital readout chain, which is based on the Caribou system~\cite{vanatCaribouVersatileData2020}, including the already developed graphical interface for controlling and measuring RD50-MPW3. Thus, it proves that configuration, readout, basic analogue behaviour and the injection circuit work as expected.

\textcolor{black}{The power consumption of the chip has been measured by monitoring the DAC currents of different supply voltages. The analogue matrix consumes 96~mW which corresponds to $\mathrm{23.4~\mu W}$ per pixel. The digital periphery consumes almost 300~mW, due to the large buffers implemented. The given values are measured in idle when fully configured.} 

\section{Conclusion and Outlook}
Based on the development of its two predecessors, the fully monolithic RD50-MPW3 chip has improvements in noise minimisation and advanced digital readout system. Preliminary measurements show full functionality of both analogue and digital circuitry. Detailed characterisation of pixel performance will be carried out in laboratory and test beam facilities
to study the tracking performance and timing behaviour of the chip. Samples will be irradiated to various fluences and tested to study the radiation tolerance of RD50-MPW3.

\acknowledgments
This work has been partly performed in the framework of the CERN-RD50 collaboration. It has received funding from the Austrian Research Promotion Agency FFG, grant number 878691, and from the European Union’s Horizon 2020 Research and Innovation program under grant agreement 101004761 (AIDAinnova).

\bibliography{RD50-MPW.bib}

\end{document}